# Astro2020 Science White Paper

# Ultra-heavy cosmic-ray science: Are r-process nuclei in the cosmic rays produced in supernovae or binary neutron star mergers?

**Thematic Areas:**

☒Formation and Evolution of Compact Objects

☒Multi-Messenger Astronomy and Astrophysics

**Principal Author:**
Name: W.R. Binns
Institution: Washington University in St. Louis
Email: wrb@wustl.edu
Phone: 314-935-6246

**Co-authors:**
M.H. Israel[1], B.F. Rauch[1], A.C. Cummings[2], A.J. Davis[2], A.W. Labrador[2], R.A. Leske[2], R.A Mewaldt[2], E.C. Stone[2], M.E. Wiedenbeck[3], T.J. Brandt[4], E.R. Christian[4], J.T. Link[4], J.W. Mitchell[4], G.A. de Nolfo[4], T.T. von Rosenvinge[4] (retired), K. Sakai[4], M. Sasaki[4], C.J. Waddington[5], H.T. Janka[6], A.L. Melott[7], G.M. Mason[8], E-S. Seo[9], J.H. Adams[10], F-K. Thielemann[11], A. Heger[12], M. Lugaro[13], A.J. Westphal[14]

**Institutions**
1. Washington University in St. Louis
2. California Institute of Technology
3. Jet Propulsion Laboratory, California Institute of Technology
4. Goddard Space Flight Center
5. University of Minnesota
6. Max Planck Institute for Astrophysics
7. University of Kansas
8. Johns Hopkins University Applied Physics Laboratory
9. University of Maryland
10. University of Alabama-Huntsville
11. University of Basel, Switzerland
12. Monash University, Melbourne, Australia
13. Konkoly Observatory, Budapest, Hungary
14. University of California, Berkely

**Executive Summary**

Galactic cosmic rays (GCR) play an important role in the dynamics of matter and magnetic fields in the interstellar medium, and probably also play an important role in star formation. In spite of their importance, and despite much recent progress, significant questions about the origin of cosmic rays remain. There is an opportunity during the coming decade to give definitive answers to these questions by making measurements of the relative abundances of every individual element, including the rarest elements with the highest atomic numbers.

The recent detection of $^{60}$Fe in the cosmic rays provides conclusive evidence that there is a recently synthesized component ($\lesssim$2.5MY) in the GCRs (Binns et al. 2016). In addition, these nuclei must have been synthesized and accelerated in supernovae near the solar system, probably in the Sco-Cen OB association subgroups ~100 pc distant from the Sun. Recent theoretical work on the production of r-process nuclei appears to indicate that it is difficult for SNe to produce the solar system abundances relative to iron of r-process elements with high atomic number (Z), including the actinides (Th, U, Np, Pu, and Cm). Instead, it is believed by many that the heaviest r-process nuclei (e.g., Just et al., 2015), or perhaps even all r-process nuclei (Siegel & Metzger, 2017), are produced in binary neutron star mergers (BNSM) (We note that recently they (Siegel, Barnes, & Metzger 2018) have argued that collapsars, which are GRBs from massive stars, may dominate r-process production). Since we now know that there is at least a component of the GCRs that has been recently synthesized and accelerated, models of r-process production by SNe and BNSM can be tested by measuring the relative abundances of these ultra-heavy r-process nuclei, and especially the actinides, since they are radioactive and provide "clocks" that give the time interval from nucleosynthesis to detection at Earth. Since BNSM are believed to be much less frequent in our galaxy than SNe (~1000 times less frequent—see Chruslinska, et al. 2018), the ratios of the actinides, each with their own half-life, will enable a clear determination of whether the heaviest r-process nuclei are synthesized in SNe or in BNSM. In addition, the r-process nuclei for the charge range $34 \leq Z \leq 82$ can be used to constrain models of r-process production in BNSM and SNe. Thus, GCRs become a multi-messenger component in the study of BNSM and SNe.

**Overview**

Recent progress in measurements of the elemental and isotopic composition of cosmic-ray nuclei with atomic number $Z < 28$ (Israel et al., 2018; Wiedenbeck, et al., 2007; Binns et al., 2005 & 2008) and the elemental abundances of the relatively rare elements with atomic number $28 < Z < 40$ (Murphy, et al. 2016; Binns, et al. 2013; Rauch, et al. 2009 ) have given strong evidence that the cosmic-ray composition is a mix of ejecta and stellar wind outflow from massive stars with old material having an elemental and isotopic composition similar to that of our Solar System. The astrophysical site for the origin of the material and acceleration is almost certainly associations of massive stars, OB associations.

Prior to the current decade, we "knew" that the neutron-rich nuclei heavier than Z=30 (r-process nuclei) in the contemporary cosmic rays were made in supernova explosions. But now, after observation of gravitational waves from neutron-star mergers, electromagnetic observations of those events, and models of r-process production in BNSM, we recognize that a substantial fraction of r-process nuclei may be made in such mergers.

It has been thought for some time that the r-process nuclear abundances are not consistent with a single r-process type. Wasserburg et al. (1996) suggested that there were "distinctive SN contributions" below and above atomic mass (A) ~140 (also see Lugaro et al. 2014). Since the advent of detailed models of BNSM (e.g. Freiburghaus, et al. 1999 and many later papers), and the much more recent detection of such a merger, it has been believed that there are likely two or more r-process production mechanisms, and some, or even all, of these are operative in BNSM. A large range of BNSM r-process production models have been developed (e.g., Siegel & Metzger, 2017; Just et al. 2015; Wu et al. 2016; Shibagaki et al. 2016; Wanajo et al., 2014). Figure 1 shows a model in which the "fission recycling peak" from BNMS is combined with "weak" and "main" peaks from SNe to produce the observed solar system r-process abundances (Shibagaki et al. 2016). Figure 2 shows a model of r-process production that includes the dynamical ejecta expelled during the neutron star merger phase and the neutrino and viscously driven outflows of the relic black hole-torus system (Just et al., 2015). BNSM models by Siegel & Metzger (2017) and Wanajo et al. (2014) can produce the full range of the solar system r-process nuclei.

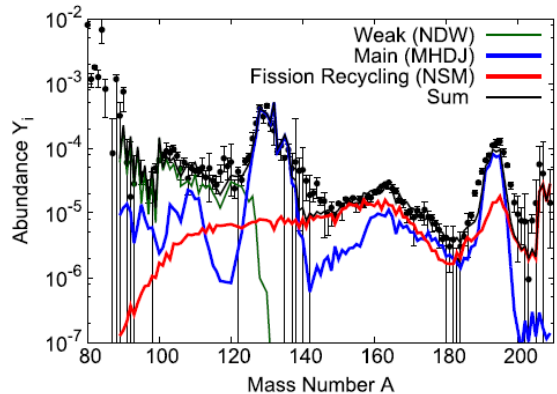

*Figure 1—Model with three different r-processes contributing to the solar-system r-process nuclei. The "weak" and "main" process come from SNe, while the "fission recycling peak" results from BNMS (Shibagaki et al. 2016).*

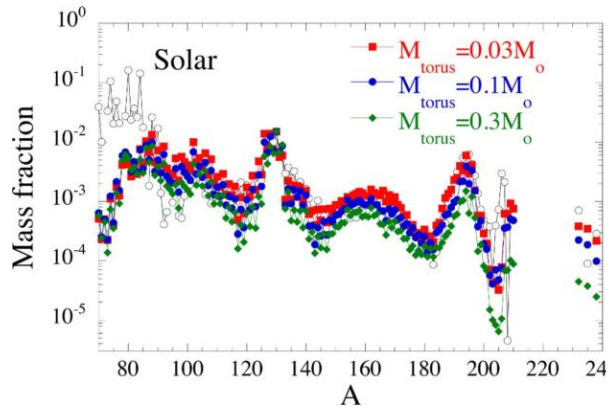

*Figure 2—Abundance distributions vs. atomic mass for black hole-torus system for different torus masses (Just et al., 2015). The open black circles are solar system r-process abundances*

In light of these observations and model results, determination of individual element abundances over the entire range of nuclei heavier than Z = 30 is particularly important. The detection of $^{60}$Fe by the Cosmic Ray Isotope Spectrometer (CRIS) on the ACE satellite is a key measurement that we can use to experimentally determine whether the heaviest r-process nuclei (the actinides) are produced by SNe or BNSM. The $^{60}$Fe cosmic-ray measurement has conclusively shown that the GCRs have at least a substantial component which has been recently synthesized near the Sun (within the last ~2.5 Myr and with distance ≲1 kpc from the Sun; Binns et al. 2016). In addition, Wallner et al. 2016 detected $^{60}$Fe in deep-sea floor sediments, and by analyzing sediment cores obtained an age of the most recent $^{60}$Fe event of ~1.5 Myr.. The $^{60}$Fe almost certainly comes from supernova ejecta, most likely from the nearby Sco-Cen OB association subgroups. If SNe are the primary contributors to the radioactive actinides ($_{90}$Th, $_{92}$U, $_{93}$Np, $_{94}$Pu $_{96}$Cm), then we should see ratios of these elements that are consistent with an age of ~1.5 Myr. Figure 3 shows the ratios of the actinides to the platinum group elements as a function of time from an r-process event. The rate of neutron star mergers in the Milky Way is estimated to be $21^{+28}_{-14}$ Myr$^{-1}$ (Chruslinkska, et al. 2017), roughly 1,000 times lower than the SN rate. (Ultra-heavy cosmic rays are lost to nuclear interactions with the interstellar medium with the result that the mean propagation distance for ~1 GeV/nucleon actinides is near 0.5 kpc (I. Moskalenko, private communication). Thus the UHCR from SNe or BNSM must be synthesized and accelerated within that volume.) If the measurements show that the mean r-process age is very large, as indicated by a lack of $_{93}$Np, $_{94}$Pu, and $_{96}$Cm, then that would point to BNSMs as the source of the heaviest r-process nuclei.

A large-area detector system placed in low-Earth orbit for a period of about three years will enable, for the first time, extension of high precision measurements of the abundance of individual elements to the rarest, heaviest elements – the actinides ($_{90}$Th, $_{92}$U, $_{93}$Np, $_{94}$Pu, and $_{96}$Cm, and perhaps beyond) There are well established instrumental techniques that can be used to realize such a detector system.

In addition to the heaviest r-process nuclei, a measurement of all predominately r-process elements with $34 \leq Z \leq 82$ can be used to provide important constraints on BNSM and SN models. Figure 4 shows an r-, s-process decomposition of solar system abundances. The key r-process elements to be measured in the 1st, 2nd, and 3rd r-process peaks are $_{34}$Se, $_{36}$Kr, $_{37}$Rb, $_{50}$Sn, $_{52}$Te, $_{54}$Xe, $_{76}$Os, $_{77}$Ir, $_{78}$Pt, and $_{82}$Pb. In addition, the s-process peaks are important for understanding the cosmic ray source.

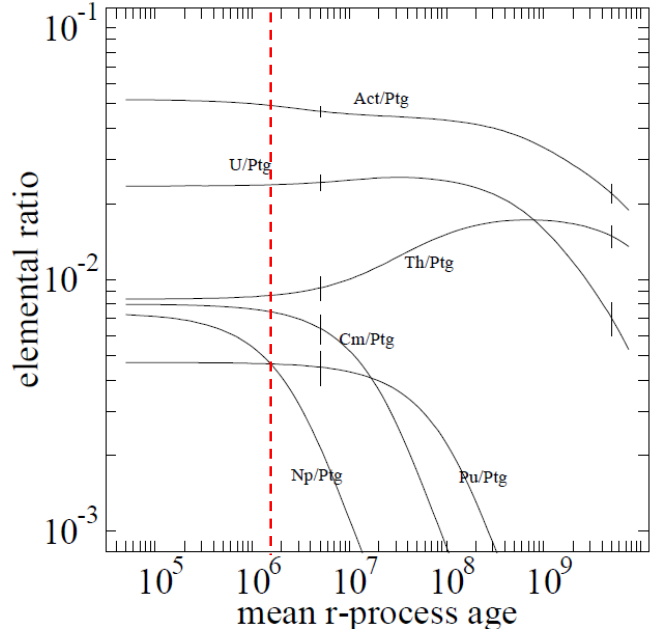

*Figure 3—The ratios of the radioactive actinides (Act) relative to platinum group nuclei (Ptg=Os+Ir+Pt)) as a function of time since an r-process event. Actinides produced in the recent, nearby SN event that produced the observed $^{60}$Fe cosmic rays should have ratios consistent with the vertical dashed red line at a mean r-process age of ~1.5 Myr. Ratios consistent with a much larger mean r-process age would point to BNSM as the source of the heaviest r-process nuclei.*

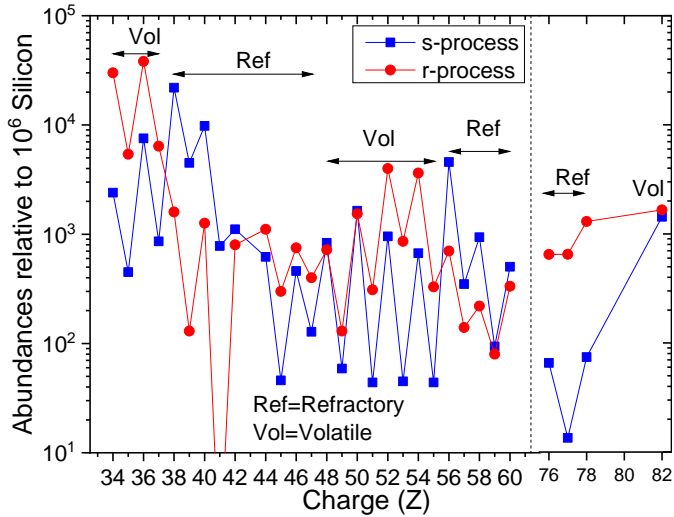

*Figure 4—r- and s-process decomposition of solar system abundances. Red symbols are predominately r-process and blue are predominately s-process elements. Refractory elements (those existing primarily in dust grains in the ISM) are indicated by the "Ref" horizontal arrows and volatile elements (those existing primarily as interstellar gas) are indicated by the "Vol" horizontal arrows.*

**Scientific Themes**

The important questions that will be addressed by such a comprehensive high-precision measurement of cosmic-ray elemental abundances are:

• Are the heaviest r-process nuclei synthesized by BNSM or by SNe?

• If BNSM are the source of the heaviest r-process nuclei, where in our galaxy is that material produced and how is it accelerated?

• Are OB associations in fact the source of the material and the location of the acceleration of most Galactic cosmic-ray nuclei?

Answers to these questions can be expected to lead to identification of the source of Galactic cosmic rays (GCR). Further, once the nature of the GCR source has been elucidated, these measurements will provide invaluable information about the local Galactic abundances of the elements, which in turn will improve our understanding of the chemical evolution of our Galaxy.

**Connections to other topics in astrophysics and nuclear physics**

The measurements described here would motivate work in theory across astronomy and physics. They would provide critical inputs to defining and constraining models of nucleosynthesis in BNSM, core-collapse SNe, and GRBs, understanding diffusive shock acceleration of cosmic rays, and improving models of GCR transport in the Galaxy. The effect of $^{60}$Fe on local cosmic rays, and consequent astrophysical and heliospheric effects of a nearby SN, are explored in detail in Frisch & Dwarkadas (2017).

There is also a strong connection with the measurements and interpretation of $^{60}$Fe deposited on the deep-ocean floor (Wallner, et al. 2016; Breitschwerdt et al. 2016), on lunar samples (Fimiani et al., 2016), and in the microfossil record (Ludwig, et al. 2016).

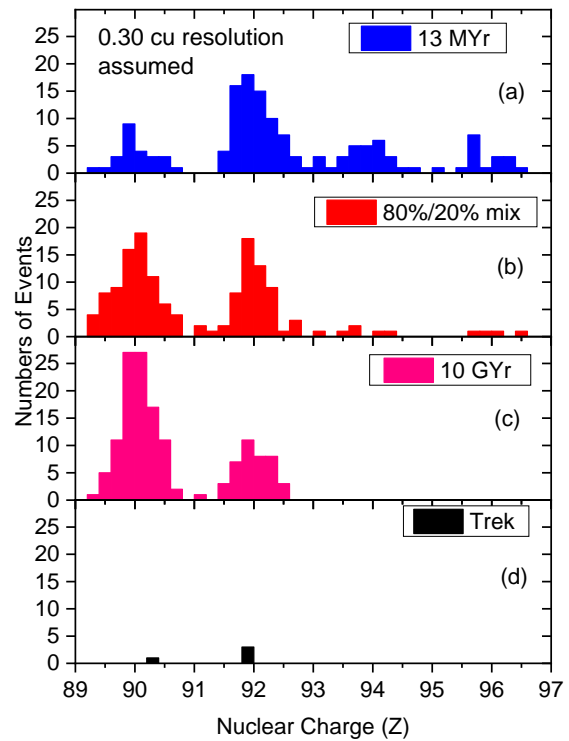

*Figure 5—Top 3 panels show expected numbers of events for different assumptions about the source material. Bottom panel shows the only cosmic ray data in the actinide range with single element resolution. (See text for details).*

**Key advances in observation and theory needed to realize opportunity**

To realize the opportunity provided by the confluence of these scientific goals and existing measurement techniques, it will be necessary to develop an instrument to be flown in space with very large effective collection power, ~30 m$^2$ sr. Such an instrument usingSiegel silicon detectors, Cherenkov detectors, and a scintillating fiber hodoscope is capable of making precision measurements of charge from neon through the actinides and beyond for energies above 0.3 GeV/nucleon. In addition, glass track detectors have been shown to have excellent resolution (Westfall, 1998). The detectors and measurement techniques used in this instrument are well understood and have extensive balloon-flight and space heritage. The space flight of an instrument with these characteristics for three years will provide the required collecting power to answer the above "central questions". Figure 5 shows a Monte Carlo calculation of the numbers of actinide events expected, assuming a charge resolution of 0.3 charge units, for different source assumptions. The top panel gives the number of events expected for a recent (13 Myr) r-process event. The next panel assumes that the cosmic-ray source is 20% massive star material (ejecta+outflow) mixed with 80% old interstellar material (ISM). The third panel is what is expected for a very old source (10 Gyr), and the bottom panel gives the total numbers of events detected to date with single element resolution (Westphal, et al. 1998). It is clear that by eye it is easy to distinguish these three assumptions for the origin of GCR source material.

The technology for accomplishing this mission is well in hand. There are no technological advances required. All that is needed is a commitment to build and launch the instrument.